\documentstyle[12pt]{article}
\def\double{\baselineskip 20pt \lineskip 20pt}

\newcommand{\be}{\begin{equation}}
\newcommand{\ee}{\end{equation}}
\newcommand{\bea}{\begin{eqnarray}}
\newcommand{\eea}{\end{eqnarray}}
\def\l{\label}

\newcommand{\pdf}{\partial}

\newcommand{\ie}{{\em i.e.\ }}

\newcommand{\Ord}{{\cal O}}
\newcommand{\Rcc}{{\cal R}}
\newcommand{\m}{m_{\rm Pl}}
\newcommand{\k}{\kappa}
\newcommand{\dd}{{\rm d}}
\newcommand{\Mpc}{{\rm\/Mpc}}
\newcommand{\MeV}{{\rm\/MeV}}

\begin{document}

\begin{titlepage}

\vspace{.2cm}
\begin{flushright}
{\small
FNAL--PUB--93/039-A\\
May 1993}
\end{flushright}
\vspace{0.1in}

\begin{center}
\Large
{\bf On Identifying the Present-day Vacuum Energy with the Potential
Driving Inflation}

\vspace{.3in}

\normalsize
{Richard A. Frewin$^1$}

\normalsize
\vspace{.2cm}
{\em Astronomy Unit, School of Mathematical Sciences, \\
Queen Mary and Westfield, Mile End Road, London E1 4NS, U. K.}
\vspace{.3cm}

\normalsize
{James E. Lidsey$^2$}

\vspace{.2cm}

\normalsize
{\em NASA/Fermilab Astrophysics Center, \\
Fermi National Accelerator Laboratory, Batavia, IL 60510, U. S. A., and  \\
Astronomy Unit, School of Mathematical Sciences, \\
Queen Mary and Westfield, Mile End Road, London E1 4NS, U. K.}

\end{center}
\baselineskip=24pt

\begin{abstract}
\noindent

There exists a growing body of observational evidence supporting a
non-vanishing cosmological constant at the present epoch. We examine
the possibility that such a term may arise directly from the potential
energy which drove an inflationary expansion of the very early
universe. To avoid arbitrary alterations in the shape of this
potential at various epochs it is necessary to introduce a
time-dependent viscosity into the system. The evolution of the
effective Planck mass in scalar-tensor theories is a natural candidate
for such an effect. In these models there are observational
constraints arising from anisotropies in the cosmic microwave
background, large-scale galactic structure, observations of the
primordial Helium abundance and solar system tests of general
relativity. Decaying power law and exponential potentials are
considered, but for these models it is very difficult to
simultaneously satisfy all of the limits.  This may have implications
for the joint evolution of the gravitational and cosmological
constants.

\vspace*{12pt}


\small e mail: $^1$raf@star.qmw.ac.uk; ~~~$^2$jim@fnas09.fnal.gov

\vspace{3cm}

To Appear {\em International Journal of Modern Physics} {\bf D}

\end{abstract}

\normalsize

\end{titlepage}

\double

\section{Introduction}
\def\theequation{\thesection.\arabic{equation}}
\setcounter{equation}{0}

The solution to vacuum general relativity with a cosmological constant
$\Lambda$ is de Sitter space and this constant and solution have often
been invoked to reconcile theory with observation. Originally Einstein
believed the universe to be static and introduced a constant
$\Lambda$-term into his field equations to cancel the expansionary
behaviour found when $\Lambda =0$.$^1$ Some decades later the steady
state scenario based on de Sitter space was developed because
observations of Hubble's constant suggested the Earth was older than
the universe itself.$^2$ More recently the inflationary scenario has
been proposed to solve some of the problems of the hot, big bang
model.$^3$ During inflation the potential energy of a quantum scalar
field dominates the energy-momentum tensor and behaves as a
cosmological constant for a finite time.

Realistic inflationary models predict that the current value of the
density parameter, $\Omega_0$, should be very close to unity.$^4$
There are a number of problems associated with the
$({\Lambda}=0,{\Omega}=1)$ universe which can be resolved if $\Lambda
\ne 0$.  Firstly, its age is $t_0 \approx 6.52h^{-1}$ Gyr, where $h$
is the current expansion rate in units of $100$ km ${\rm sec}^{-1}$
${\rm Mpc}^{-1}$.  This is very close to the age of the oldest
globular clusters in the galaxy, $t_{\rm GC} =(13-15)\pm 3$ Gyr, if $h
\ge 0.6$ as suggested by some observations.$^5$ If ${\Lambda} \ne 0$,
however, the expansion rate is increased and $t_0$ may exceed $t_{\rm
GC}$ if ${\Omega}_{\rm vacuum} \approx 0.8$.$^6$ Moreover, most
dynamical determinations of $\Omega_0$ suggest $\Omega_0 = 0.2 \pm
0.1$ up to scales 30 Mpc and the apparent inconsistency with inflation
is again resolved if $\Omega_{\rm vacuum} \approx 0.8$.$^7$ Finally,
the introduction of vacuum energy into the standard cold dark matter
model of galaxy formation accounts for the extra large-scale
clustering observed in galaxy surveys.$^8$

Hence, there are a number of reasons for supposing that a cosmological
constant may be influential at the present epoch.$^9$ This work
investigates whether the potential energy that drove the inflationary
expansion could be such a term.  This has been investigated previously
within the context of general relativity by Peebles and Ratra,$^{10}$
but their models required the form of the potential to change
drastically at various epochs and therefore suffered from an element
of ad-hoc fine-tuning. If the potential is relevant at the current
epoch, it must either have a minimum at $V\ne 0$ or contain a
non-vanishing decaying tail. Although the first possibility is not
ruled out, it requires severe fine\---tuning, so we focus on the
second. This implies that thermalization of the false vacuum will not
proceed via rapid oscillations of the scalar field about some global
minimum and the form of the potential must change at various epochs.
In general relativity the potential must be sufficiently flat for the
strong energy condition to be initially violated, but must then become
steep enough for reheating to proceed. But the energy density of the
field must redshift at a slower rate than the ordinary matter
components at late times if the vacuum energy is to once more dominate
the dynamics.$^{54}$

Instead of altering the shape of the potential we extend the
gravitational sector of the theory beyond general relativity and
investigate whether inflation was driven by potentials which are (a)
too steep to lead to inflation in general relativity and (b) do not
contain a global minimum. We shall refer to these as {\em steep\/}
potentials. A number of unified field theories lead to scalar field
potentials which exhibit both of these characteristics. The mechanism
leading to inflation is very simple.  In pure Einstein gravity
containing a single, minimally coupled scalar {\em inflaton} field,
$\sigma$, the strong energy condition is violated if the condition
${\dot{\sigma}}^2<V$ holds, \ie the potential energy dominates over
the field's kinetic energy.  Clearly this condition must break down at
some point as steeper potentials are considered. But a finite interval
of inflation is possible with steep potentials if one introduces a
viscosity into the inflaton field equation which decays as the
universe expands.  This viscosity slows the field down and can lead to
inflation. As the viscosity becomes weaker, however, the inflaton's
kinetic energy increases significantly and a natural exit from
inflation proceeds as the expansion becomes subluminal. We identify
the dilaton field which arises in scalar-tensor theories as the
natural source of this viscosity.

This paper is organised as follows.  We survey theories that lead to
inflation with steep potentials in section 2.  In section 3 we derive
expressions for the amplitudes of the primordial fluctuation spectra
and discuss the most stringent observational constraints which any
viable model of this type must satisfy. These limits arise from the
observations of large\---scale galactic structure,$^{11}$ anisotropies
in the cosmic microwave background radiation (CMBR),$^{12}$
nucleosynthesis calculations$^{14,15}$ and time\---delay experiments
in the solar system.$^{16}$ Numerical results for both decaying power
law and exponential potentials are presented in section 4 and we
conclude that successful inflation based on this mechanism is unlikely
for the examples considered. Some possible implications of this
conclusion are discussed in section 5.  Unless otherwise stated, units
are chosen such that $c=\hbar=1$, and the present day value of the
Planck mass is normalized to $\m =1$.

\section{Inflation with steep potentials}
\def\theequation{\thesection.\arabic{equation}}
\setcounter{equation}{0}

\subsection{Scalar-Tensor theories}

A suitable source of the viscosity is the dilaton field which arises
in the Bergmann-Wagoner class of theories.$^{17}$ This field plays the
role of a time-dependent gravitational constant.  The field equations
for these scalar-tensor theories are derived by varying the action
\be
\l{eq2.1}
S= \int \dd^4x {\sqrt{-g}} \left[ h(\psi){\cal R} -{1 \over 2}
({\nabla\psi})^2-
U(\psi) -16{\pi} \left( {1 \over 2} ({\nabla\sigma})^2 +V(\sigma) + {\cal
L}_{\rm matter} \right) \right] ,
\ee
where $g={\rm det} g_{\mu\nu}$, ${\cal R}$ is the Ricci scalar,
$h(\psi)$ is some arbitrary function of the dilaton $\psi$, $U(\psi)$
is the dilaton self-interaction and ${\cal L}_{\rm matter}$ is the
Lagrangian of a perfect baryotropic fluid, which we assume to be
relativistic matter with an effective equation of state
$p_r={\rho}_r/3$.  The dilaton and inflaton field equations are
coupled and the extra viscosity on ${\sigma}$ is due to the dynamical
evolution of the effective Planck mass in the theory. The strength of
gravity is determined by the magnitude of $h(\psi)$ and a fraction of
the inflaton's potential energy is converted into the dilaton's
kinetic energy rather than contributing to ${\dot{\sigma}}^2$.

It is well known that theories of this type appear as the low energy
limits to a number of unified field theories and can be expressed in
the Jordan-Brans-Dicke (JBD) form with a variable ${\omega}$-parameter
by defining a new scalar field ${\Psi} \equiv h(\psi)$.$^{18}$ The
action (\ref{eq2.1}) becomes
\be\l{eq2.2}
S=\int \dd^4x{\sqrt{-g}} \left[ {\Psi}{\cal R}-{{{\omega}(\Psi)} \over {\Psi}}
({\nabla\Psi})^2-U(\Psi)-16{\pi} \left(  {1 \over
2}({\nabla\sigma})^2+V(\sigma) +{\cal L}_{\rm matter} \right) \right]
\ee
where
\be\l{eq2.4}
\omega(\Psi)=\frac{h(\psi)}{2(\dd h/\dd{\psi})^2}.
\ee
The theory of general relativity is recovered whenever a turning point
exists in the functional form of $h(\psi)$ (see section 3.3), because
${\omega}(\Psi)$ tends to infinity and the dilaton's kinetic energy
decouples. Although the theories (\ref{eq2.1}) and (\ref{eq2.2}) are
equivalent mathematically, there is a philosophical difference which
arises in deciding which function $h(\psi)$ or ${\omega}(\Psi)$ should
be treated as the fundamental quantity.$^{19}$ In general, if we
consider a simple form for $h(\psi)$ such as a truncated Taylor
series, this leads to a very complicated form for ${\omega}(\Psi)$ and
vice-versa.

For example  we can expand $h(\psi)$ as some power series
\be
\l{eq2.5}
h(\psi) =\sum^{\infty}_{i=0} {\alpha}_i{\psi}^i
\ee
for arbitrary constant coefficients ${\alpha}_i$. To lowest order  $h(\psi)
\approx {\alpha}_0$, but this corresponds to a `constant' gravitational
constant and is not interesting.  Moreover the linear term may always
be eliminated by a simple field redefinition, so the lowest order of
interest is the quadratic contribution and this is simply the standard
JBD theory with constant $\omega(\Psi)$.  In principle, given suitable
initial conditions, inflation will then occur as ${\sigma}$ slowly
rolls down its potential causing ${\psi}$ to increase.  Eventually
higher-order $(i>2)$ terms in the expansion (\ref{eq2.5}) will become
important and for appropriate choices of ${\alpha}_i$, such as
${\alpha}_2>0$ and ${\alpha}_3<0$, one can easily arrange for a local
maximum in $h(\psi)$ to exist at some value ${\psi}={\psi}_0$. At this
point the theory will be identical to general relativity if we
normalise $h({\psi}_0)=1$.  It is clear that as ${\psi}$ approaches
${\psi}_0$, inflation will end because the gravitational friction
weakens and the inflaton speeds up. This model is a chaotic version of
the hyperextended scenario and proceeds via a second-order phase
transition.$^{18}$

Another possibility is the induced theory of gravity where $h(\psi)$
is quadratic but the dilaton potential is non-vanishing and contains a
global minimum at ${\psi}_0$. Inflation driven by ${\sigma}$ could
occur if ${\psi}$ is initially displaced from ${\psi}_0$, but will
clearly end as Einstein gravity is recovered and the dilaton settles
into this minimum. After spontaneous compactification to
four-dimensions, some Kaluza-Klein theories have this structure, where
the dilaton is identified as the logarithm of the radius of the
internal space.$^{20}$ If monopole and Casimir effects due to
non-trivial field configurations are also considered, a classically
stable ground state is possible at ${\psi}_0$.$^{21}$

Theories (\ref{eq2.1}) and (\ref{eq2.2}) are conformally equivalent to
general relativity with a matter sector containing two interacting
scalar fields.$^{22}$ By redefining the graviton and dilaton fields as
\be
\l{eq2.6}
{\tilde{g}}_{\mu\nu}={\Omega}^2g_{\mu\nu}, \qquad {\Omega}^2
\equiv 2{\kappa}^2h(\psi)
\ee
and
\be
\l{eq2.7}
{\kappa\Phi}= \int \dd{\psi} \left( {{{3(\dd h/\dd{\psi})^2+h(\psi)} \over
{2h^2(\psi)}}} \right)^{1/2},
\ee
where ${\kappa}^2=8{\pi}\m^{-2}$,  the action (2.1) for $U(\psi)=0$ may be
rewritten in the Einstein-Hilbert form
\be
\l{eq2.8}
S= \int \dd^4x {\sqrt{-{\tilde{g}}}} \left[ {{\tilde{R}} \over
{2{\kappa}^2}}-{1
\over 2} ({\tilde{\nabla}}{\Phi})^2-{1 \over 2}A(\Phi)({\tilde{\nabla}}
{\sigma})^2-C({\Phi})V({\sigma}) \right],
\ee
where
\be
\l{eq2.9}
A^{-1}(\Phi) \equiv {\Omega}^2 =2{\kappa}^2h(\psi), \qquad C(\Phi) \equiv
A^{2}(\Phi)
\ee
and we assume that $\{h,\dd h/\dd{\psi}\}>0$ for consistency. The
viscosity appears through the non-standard coupling, $A(\Phi)$, of the
${\Phi}$-field to the inflaton's kinetic term. This coupling evolves
towards unity as ${\Phi}$ settles into a minimum of the effective
potential $C(\Phi)V(\sigma)$, thus causing the accelerated expansion
to end. In principle one can therefore realise this scenario in
general relativity if the matter sector of the theory is modified in
an appropriate fashion. Some higher\---dimensional, higher\---order
gravity theories exhibit this conformal structure upon
compactification of the extra dimensions.$^{23}$

In the following discussions we refer to $g_{\mu\nu}$ as the
Jordan-Brans-Dicke (JBD) frame and ${\tilde{g}}_{\mu\nu}$ as the
Einstein-Hilbert (EH) frame. In the former it is the matter
contributions which are canonical, whereas gravity is canonical in the
latter.  The evolution of the Planck mass in the JBD frame is
translated into a time-dependence for gauge and Yukawa couplings in
the EH frame.$^{24}$

\subsection{Field Equations}

Extremizing the action (\ref{eq2.2}) with respect to arbitrary
variations of the metric produces the gravitational field equations
\bea\label{GravFE}
R_{\mu\nu}-\frac{1}{2}g_{\mu\nu}\Rcc & = & -
\frac{1}{2}g_{\mu\nu}\frac{U}{\Psi} -
\frac{8\pi}{\Psi}T_{\mu\nu} - \frac{\omega(\Psi)}{\Psi^{2}}
\left[\nabla_{\mu}\Psi\nabla_{\nu}\Psi - \frac{1}{2}g_{\mu\nu}g^{\rho\lambda}
\nabla_{\rho}\Psi\nabla_{\lambda}\Psi\right] \nonumber\\
\mbox{} & - & \frac{1}{\Psi}
\left[\nabla_{\mu}\nabla_{\nu}\Psi - g_{\mu\nu}\Box\Psi\right],
\eea
where $\Box\equiv g^{\mu\nu}\nabla_{\mu}\nabla_{\nu}$ and the
energy-momentum tensor $T_{\mu\nu}$ is defined as the functional
derivative of the matter lagrangian including the inflaton
field.$^{25}$ To proceed we assume a space-time with isotropic and
homogeneous spatial sections (the Friedmann-Robertson-Walker
universes) where the fields are functions of cosmic time only. In this
case minimising (\ref{eq2.2}) with respect to variations in $\sigma$
and $\Psi$ leads to the equations
\be
\l{ife}
\ddot{\sigma} + 3H\dot{\sigma} = - \frac{\dd V}{\dd\sigma}
\ee
and
\be\label{Ricci}
\Rcc = 2\omega\left(\frac{\ddot{\Psi}}{\Psi} +
3H\frac{\dot{\Psi}}{\Psi}\right) - \omega
\left(\frac{\dot{\Psi}}{\Psi}\right)^{2} +
\frac{\dd\omega}{\dd\Psi}\frac{\dot{\Psi}^{2}}{\Psi} +
\frac{\dd U}{\dd\Psi},
\ee
respectively. Combining Eq. (\ref{Ricci}) with the trace of (\ref{GravFE}) we
arrive at the dilaton field equation
\be\label{dilaton}
(2\omega + 3)\left(\ddot{\Psi} + 3H\dot{\Psi}\right) = 2U -
\Psi\frac{\dd U}{\dd\Psi} - \frac{\dd\omega}{\dd\Psi}\dot{\Psi}^{2} + 8\pi T,
\ee
where $T=\rho_{total}-3p_{total}$ is the trace of $T_{\mu\nu}$.   Finally,
the time-time component of (\ref{GravFE}) gives the Friedmann equation
\be
3\left(H^{2}+\frac{k}{a^{2}}\right)=
\frac{\omega}{2}\left(\frac{\dot{\Psi}}{\Psi}\right)^2 +
\frac{1}{2}\frac{U}{\Psi} - 3H\frac{\dot{\Psi}}{\Psi} +
\frac{8\pi}{\Psi}\left(\frac{1}{2}\dot{\sigma}^{2}+V(\sigma)+
\rho_{r}\right).
\ee
where $k=-1,0,+1$ determines the spatial curvature.

It is necessary to reheat the universe shortly after the expansion has
become subluminal. The couplings of the inflaton to other relativistic
matter fields should then become important because the potential is
steep.  However, reheating in these theories is difficult to model so
we follow Morikawa \& Sasaki$^{26}$ by introducing a phenomenological
dissipation term into the field equations.  It was shown that for an
exponential inflaton potential and a Yukawa type coupling the inflaton
field equation (\ref{ife}) becomes$^{27}$
\be
\ddot{\sigma} + 3H\dot{\sigma}  = - \frac{\dd V}{\dd\sigma}
- C_{v}\dot{\sigma},
\ee
where to zeroth order the dissipation
\be
C_{v}\sim fM_{\sigma}=f\sqrt{\frac{\dd^{2}V}{\dd\sigma^{2}}}
\ee
is proportional to the effective mass of the inflaton and $f=\Ord(1)$.
This form for $C_{v}$ should be appropriate for any steep potential.
The Bianchi identity $\nabla_\mu T^{\mu\nu} \equiv 0$ then implies
that
\be
\dot{\rho}_{r} = - 4H\rho_{r} + C_{v}\dot{\sigma}^{2},
\ee
where $\rho_{r}$ is the energy density of relativistic particles. The
decay product must lead to baryogenesis, and we assume the reheat
temperature is significantly larger than its rest mass $m_X$. We may
then treat it as a relativistic fluid in local thermodynamic
equilibrium. Once the inflaton's effective mass falls below $m_X$ the
dissipation becomes negligible.

\subsection{Change of Independent Variable}

Numerical solutions must be found which trace the evolution of the
universe from the Planck time to the current epoch. This corresponds
to some $60$ orders of magnitude in the independent variable. The
codes used to produce these solutions become inefficient when
integrating over more than $15$ orders of magnitude. Therefore a
change of independent variable was used to speed up the process of
generating solutions.

The number of e-foldings $N=\ln a$ was used as the new independent
variable since this is clearly a monotonically varying function for
expanding universes.  From the relation $H=\dot{a}/a=\dd(\ln a)/\dd t$
it can be seen that $\dd t=\dd N/H$.  Hence for a variable $x(t)$ we
must make the substitutions: $\dot{x}\rightarrow Hx'$ and
$\ddot{x}\rightarrow H^{2}x'' + HH'x'$ where a prime denotes
differentiation with respect to $N$.  We now have the following
equations for $k=0$:
\be\l{infmotN}
\sigma'' + \left(3 + \frac{HH'}{H^{2}}\right)\sigma' +
\frac{1}{H^{2}}\frac{\dd V}{\dd\sigma} + \frac{C_{v}}{H}\sigma' = 0 \\
\ee\be
\rho'_{r} + \rho_{r} - HC_{v}\sigma'^{2} = 0 \\
\ee\be
(2\omega+3)\left[\Psi'' + \left(3 + \frac{HH'}{H^{2}}\right)
\Psi'\right] - 2\frac{U}{H^{2}} + \frac{\Psi}{H^{2}}\frac{\dd U}{\dd\Psi} +
\frac{\dd\omega}{\dd\Psi}(\Psi')^{2} = 8\pi\left[\frac{4V(\sigma)}{H^{2}} -
\sigma'^{2}\right] \\
\ee\be
t=\int\frac{\dd N}{H} \\
\ee
and
\be\l{fredN}
H^{2}= \left[ \frac{1}{2}\frac{U}{\Psi} +
\frac{8\pi}{\Psi}\left(V+\rho_{r}\right) \right]
\left[ 3-\frac{\omega}{2}\left(\frac{\Psi'}{\Psi}\right)^{2} +
3\frac{\Psi'}{\Psi} + \frac{8\pi}{2}\frac{\sigma'^{2}}{\Psi} \right]^{-1}.
\ee

We now need an extra equation to calculate $H'$.  This is found
by taking (\ref{Ricci}) and substituting for
$(\ddot{\Psi}+3H\dot{\Psi})$ from (\ref{dilaton}) leaving
\bea
6\dot{H}+12H^{2} & = & \left(1-\frac{2\omega}{2\omega+3}\right)\left(
\frac{\dd\omega}{\dd\Psi}\frac{\dot{\Psi}^{2}}{\Psi} +
\frac{\dd U}{\dd\Psi}\right)
\nonumber\\ \mbox{} & + &
\frac{2\omega}{2\omega+3}\frac{1}{\Psi}\left(2U+8\pi\left(4V(\sigma)-
\dot{\sigma}^{2}\right)\right) -
\omega\left(\frac{\dot{\Psi}}{\Psi}\right)^{2},
\eea or converting to independent variable $N$:
\bea\l{rayN}
HH' & = & \frac{1}{6}\left(1-\frac{2\omega}{2\omega+3}\right)\left(
H^{2}\frac{\dd\omega}{\dd\Psi}\frac{(\Psi')^{2}}{\Psi} +
\frac{\dd U}{\dd\Psi}\right) \nonumber\\
\mbox{} & + & \frac{1}{6}\frac{2\omega}{2\omega+3}
\frac{1}{\Psi}\left(2U+8\pi\left(4V(\sigma) - H^{2}\sigma'^{2}\right)\right)
 \nonumber\\ \mbox{} & - & H^{2}\omega\left(\frac{\Psi'}{\Psi}\right)^{2}
-2H^{2}.
\eea

Defining $y=\sigma'$ and $z=\Psi'$ we obtain a set of first order, ordinary
differential equations which can be integrated given initial conditions on
$\{t,\rho_{r},\sigma,\sigma',\Psi,\Psi'\}$ at some initial $N=N_{i}$. When
certain
slowroll conditions, such as $\ddot{\Psi} \ll H\dot{\Psi}$ and $\dot{\sigma}^2
\ll
V$ are valid, this coordinate transformation will simplify any
analytical approach considerably.

Before proceeding to solve these equations for specific inflaton
potentials, we shall investigate the most important observational
constraints that any successful scenario of this type must satisfy. We
present a detailed discussion of how such limits arise since they will
always be relevant in any future analysis.

\section{Constraints from the CMBR, nucleosynthesis and the solar system}
\def\theequation{\thesection.\arabic{equation}}
\setcounter{equation}{0}

\subsection{Amplitudes of scalar and tensor perturbations in the conformal
frame}

The purpose of this section is to derive the general formulae for the
amplitude of density fluctuations which arise in models based on
theory (\ref{eq2.1}) and then discuss the constraints which any
scalar\---tensor scenario based on a steep potential must satisfy.  We
generalize the calculation of Berkin and Maeda$^{28}$ by considering
an arbitrary functional form for $h(\psi)$ or equivalently for
${\omega}(\Psi)$.  If one wishes to employ Bardeen's$^{29}$ analysis
for the evolution of super-horizon sized perturbations, it is
necessary for consistency to perform the calculation in the EH frame
where the Planck mass is truly constant. To proceed analytically,
however, it is also necessary to assume certain `slow-roll'
approximations for the two fields and ignore all quadratic
first-derivative and linear second-derivative terms in the field
equations. The full equations derived from theory (\ref{eq2.8}) were
presented in Ref. 30 and for slowly rolling fields they reduce to
\bea
\l{eq3.1a}
{\tilde{H}^2} \approx {{{\kappa}^2} \over 3} CV \\
\l{eq3.1b}
3{\tilde H}{\dd_{\eta}{\Phi}} \approx -C_{\Phi}V \\
\l{eq3.1c}
3{\tilde{H}}{\dd_{\eta}{\sigma}} \approx - A V_{\sigma},
\eea
where subscripts ${\Phi}$ and ${\sigma}$ denote differentiation with
respect to ${\Phi}$ and ${\sigma}$ respectively, $\dd_{\eta} \equiv
\dd/\dd{\eta}$, ${\eta} \equiv \int\dd t{\Omega}(t)$ denotes cosmic
time in the conformal picture and tildes refer to quantities defined
in the EH frame.  In general, the total energy density of the system
is defined by
\be
\l{eq3.2}
{\tilde{\rho}} \equiv {1 \over 2}({\dd_{\eta}{\Phi})^2}+{1 \over
2}A(\dd_{\eta}{\sigma})^2+A^2V.
\ee
and the expression for the density spectrum can be found by extending
the results of Lyth$^{31}$ to the case of two interacting scalar
fields.  If we denote by ${\eta}_1$ and ${\eta}_2$ the times a
perturbation leaves and re-enters the horizon in the EH frame, then
\be
\l{eq3.3}
\left[(1+{\beta}) {{{\delta}{\tilde{\rho}} \over
{\tilde{\rho}}}} \right]_{{\eta}_1}= \left[ (1 +{\beta})
 {{{\delta{\tilde{\rho}}}} \over {{\tilde{\rho}}}}
\right]_{{\eta}_2}
\ee
where ${\beta}=2/[3(1+{\omega})]$ and ${\omega} ={\tilde
p}/{\tilde{\rho}}$ is the ratio of the pressure to the energy density
at the two epochs. In this discussion, the term `horizon' refers to
the inverse Hubble scale at the times ${\eta}_i$ and horizon crossing
is defined in terms of comoving wavenumber, ${\tilde{k}}(\eta)$, by
the expression ${\tilde{k}}(\eta)={\tilde a}(\eta){\tilde H}(\eta)$.

During inflation, ${\omega}_1 \approx -1$ which implies
\be
\l{eq3.4}
\left[{{{\delta}{\tilde{\rho}}} \over {\tilde{\rho}}}
\right]_{{\eta}_2} \approx {2 \over 3} \left( {1 \over {1+{\beta}_2}}
\right)  \left[ {{{\delta}{\tilde{\rho}}} \over {{\tilde p}
+{\tilde{\rho}}}}  \right]_{{\eta}_1}.
\ee
It is necessary to derive an expression for ${\delta}{\tilde{\rho}}$
at the first horizon crossing. This is achieved by varying
(\ref{eq3.2}) and ignoring quadratic terms in ${\dd_{\eta}{\Phi}}$ and
${\dd_{\eta}{\sigma}}$.  Dimensional analysis implies
${\delta}({\dd_{\eta}{\Phi}})
\approx {\tilde H}({\delta}{\Phi})$ and ${\delta}(\dd_{\eta}{\sigma}) \approx
{\tilde H} {\delta\sigma}$, so
\be
\l{eq3.5}
{\delta}{\tilde{\rho}} \approx -2({\tilde H}{\dd_{\eta}{\Phi}} {\delta\Phi}
+A(\Phi){\tilde H}{\dd_{\eta}{\sigma}}{\delta\sigma}),
\ee
where the field equations (3.1) have been used to simplify the
result. The terms ${\delta\Phi}$ and ${\delta\sigma}$ are stochastic
quantities arising from quantum fluctuations in the fields. In
practise, one uses the two-point correlation functions to estimate
their magnitudes.$^{32}$ However, in this theory the inflaton has a
non-standard kinetic term in the action (\ref{eq2.8}). Naively, one
would expect $|{\delta\sigma}| \approx {\tilde H}$, but an additional
factor is present in this expression which arises when the second
quantization is performed. In general, this factor is given by the
inverse square root of the function coupled to the kinetic term of the
inflaton. Hence
\be
\l{eq3.6}
|{\delta\Phi}| \approx {\tilde H}, \qquad |{\delta\sigma}|
\approx A^{-1/2}(\Phi){\tilde H}.
\ee
By substituting (\ref{eq3.6}) into (\ref{eq3.4}), we arrive at the final result
\be
\l{eq3.7}
{\tilde{A}}_S \equiv \left[{{{\delta}{\tilde{\rho}}} \over {\tilde{\rho}}}
\right]_{{\eta}_2}  \approx {\alpha} {\tilde{H^2}}  \left[{{|{\dd_{\eta}
{\Phi}}| + {\sqrt{A}}|{\dd_{\eta}{\sigma}}| } \over
{{(\dd_{\eta}{\Phi})^2+A(\dd_{\eta}{\sigma})^2}}}
\right]_{{\eta}_1 \approx {\eta}_{60}}
\ee
where ${\alpha}$ is a numerical constant of order unity and the right-hand
side is evaluated at the start of the last $60$ e-folds of inflation. This
is a very general expression and is valid for arbitrary functions $\{h(\psi),
V(\sigma) \}$.

Two regions may be defined in terms of the relative values of
${\dd_{\eta}{\Phi}}$ and ${\sqrt{A}}{\dd_{\eta}{\sigma}}$ as
\bea
{\rm Region \quad I} \quad - \quad |{d_{\eta}{\Phi}}|< {\sqrt{A}}
|{\dd_{\eta}{\sigma}}| \\ {\rm Region \quad II} \quad - \quad
|{\dd_{\eta}{\Phi}}|
> {\sqrt{A}}|{\dd_{\eta}{\sigma}}|.
\eea
In these two limiting cases, the amplitude of density  perturbations becomes
\bea
\l{eq3.14}
{\tilde{A}}_{S,{\rm I}} \approx {{{\alpha}{\kappa}^3} \over {\sqrt{3}}} A^{3/2}
{{V^{3/2}} \over {V_{\sigma}}}\\
\l{eq3.15}
{\tilde{A}_{S,{\rm II}}} \approx {{{\alpha}{\kappa}^3} \over {\sqrt{3}}}
{{C^{3/2}} \over {C_{\Phi}}} V^{1/2},
\eea
where the field equations (3.1) have been used. In region II, the evolution of
the
dilaton field, as represented by the ${\Phi}$-field, dominates  the
inflationary dynamics and one recovers the results for the original
hyperextended
scenario.$^{18}$

The general formula for the perturbation spectrum in the EH frame can
be expressed in terms of quantities in the JBD frame with the use of
equations (\ref{eq2.6}) and (\ref{eq2.7}). The conformal
transformation (\ref{eq2.6}) implies that the two-point correlation
function of the inflaton in the JBD frame is $|{\delta\sigma}| \approx
H$, as expected. It is straightforward to show that the expression for
$\tilde{A}_S$ is given by
\be
\l{eq3.16}
{\tilde{A}_S} \approx  {\alpha}A^{1/2} H^2 \left[
{|{{\dot{\Phi}}|+A^{1/2}|{\dot{\sigma}}|} \over
{{{\dot{\Phi}}^2+A{\dot{\sigma}}^2}}} \right].
\ee
The results of McDonald$^{33}$ were derived in the JBD frame and they can be
compared to those of  Ref. 28 for the special case
of the JBD theory  by using  Eq. (\ref{eq2.6}). For the
pure JBD theory, Eq. (\ref{eq3.16}) reduces to
\be
\l{eq3.17}
{\tilde{A}}_S  \approx {\alpha}H^2
\left[ {{(1+6{\epsilon})^{1/2}|{\dot{\psi}}|+|{\dot{\sigma}}|} \over
{(1+6{\epsilon}){\dot{\psi}}^2+{\dot{\sigma}}^2}} \right],
\ee
where $\epsilon \equiv 1/4\omega$ and this is identical to Eq. (47) in
McDonald's paper  with the
translation
\be
\l{eq3.18}
\dd/\dd t \quad \longrightarrow \quad (1+6{\epsilon})^{1/2} \dd/\dd t
\ee
for the time derivative of the dilaton.  Such a change is very close
to unity when ${\epsilon} \ll 1$. Hence the change in Newton's
constant is negligible compared to the evolution of the perturbations,
as assumed by McDonald. (When ${\epsilon} \ge 1/6$, the slowroll
conditions used to derive (3.1) break down and the derivation of
Eq. (\ref{eq3.7}) is inconsistent).

The close agreement between the expressions for the density spectrum
in the two frames suggests that the conformal transformation
(\ref{eq2.6}) used to recast the JBD theory into the Einstein-Hilbert
form may be valid at the semi-classical level in these chaotic
models. The two results should be identical in the limit as
${\epsilon} \rightarrow 0$. A detailed comparison of the two different
methods was made by Guth and Jain$^{34}$ within the context of old
extended inflation in which the inflaton is fixed.  These authors
extended a previous analysis by Kolb {\it et al.}$^{35}$ They also
conclude that the technique of conformal transformations is valid up
to a numerical factor of order unity when ${\epsilon} \ll 1$.

Finally the expression for the amplitude of gravitational waves
(tensor modes) should also be calculated in the EH frame. One can view
a graviton as a minimally coupled massless scalar field with two
degrees of freedom corresponding to the two polarization states of the
wave. Abbott and Wise$^{36}$ have derived an expression for the
amplitude at horizon crossing for an arbitrary inflationary solution,
so their results are also valid for the theories under consideration
in this work. Therefore, to a first approximation, the equivalent
expression for the tensor modes is
\be
\l{eq3.19}
{\tilde{A}}_G={{\kappa} \over {4{\pi}^{3/2}}}{\tilde H}.
\ee

To summarize, we have derived the expressions for the scalar
$({\tilde{A}}_S)$ and tensor $({\tilde{A}}_G)$ modes in the EH
frame. For comparison with observation, however, we require the
equivalent expressions in the JBD frame, since we are interpreting
this as the physical frame.  In general, the equivalent expressions
$A_S$ and $A_G$ are related to their counterparts in the EH frame by
an expression involving the conformal transformation
(\ref{eq2.6}). The relation therefore depends on the form of
$h(\psi)$.  However, if the condition
\be
\l{eq3.20}
{{\dd_{\eta}{\Omega}} \over {{\Omega}{\tilde H}}} \ll 1
\ee
holds, and if ${\Omega}$ is a function of $t$ only, it is straightforward
to show that the scale factors and expansion rates  in the two universes are
related by
\be
\l{eq3.21}
a(t)={\Omega}^{-1}(t){\tilde a}(\eta), \qquad H={\Omega}{\tilde H}.
\ee
Eq. (\ref{eq3.21}) implies that ${\tilde k}={\tilde a}{\tilde H}
\approx aH =k$, so the scale dependences in the two frames are
approximately equal. Moreover, the definition of the energy density,  ${\rho}
\equiv g^{00}T_{00}$, implies
\be
\l{eq3.22}
{{{\delta}{\tilde{\rho}}} \over {{\tilde{\rho}}}}={{\delta\rho} \over
{\rho}}-4{{\delta\Omega} \over {\Omega}}
\ee
and we conclude that amplitudes are also equal when ${\delta\Omega}/{\Omega}
\ll 1$, as implied by Eq. (\ref{eq3.20}) for $\dd{\eta} \approx
{\tilde{H}}^{-1}$.$^{37}$
Therefore, it is sufficient to consider results from the conformal (EH) frame
directly.

\subsection{The CMBR and large-scale structure}

If particle physics specified the unique inflaton potential, the
amplitude $A_S$ would be known for all scales. Because there exist
many possible models, however, the amplitude of fluctuations must be
normalized using observations of large-scale structure.  The CfA
survey suggests that the {\em rms\/} fluctuation in the galaxy counts
is unity in a sphere of radius $8h^{-1}\Mpc$ and we normalize the
amplitude by specifying the {\em rms\/} fluctuation in the mass
distribution to be ${\sigma}_8=b_8^{-1}$ within this sphere.$^{53}$
$b_8$ is the biasing factor which we assume to be constant over the
scales of interest.

We shall show in section (5) that the inflationary solutions
approximately 60 e-foldings before reheating can be expressed as a
power law $a(t) \propto t^p$ for $p \gg 1$. The slowroll
approximations are valid in this region of parameter space.  It is
well known that this solution leads to a primordial power law spectrum
of scalar fluctuations, $P(k)
\propto A_S^2(k)k \propto k^n$, where $n=1-2/(p-1)$ is the spectral index.
The spectrum of tensor modes has an identical scale-dependence for this
solution  when $n \le 1$, \ie $A_G \propto A_S$, and scale-invariant
fluctuations correspond to $n=1$ $(p= \infty )$.$^{38}$

Equations relating ${\sigma}_8$ to $n$ have been derived using results
from the CMBR and large-scale structure.$^{12}$ The range of parameter
space $({\sigma}_8,n)$ consistent with these observations can then be
determined.$^{39}$ For our purposes, these constraints restrict the
value of the effective JBD parameter (\ref{eq2.4}) during inflation.

At present the observation of multipole anisotropies by the {\em
COBE\/} DMR experiment provides the strongest constraint from the
CMBR. The root of the variance at $10^o$ is observed to be
${\sigma}_T(10^o)=(1.1 \pm 0.2) \times 10^{-5}$, where the mean
blackbody temperature $T_0=2.736$ K is taken and limits are to
1-sigma.$^{12}$ On $10^o$ scales reionization processes such as
Thompson scattering are not important and the dominant contribution to
the anisotropy is from the Sachs-Wolfe effect when photons at
decoupling are redshifted as they climb out of potential wells.$^{41}$
For power law inflation the observed anisotropy is due to both scalar
{\em and} tensor modes.$^{42}$ Following an identical procedure to
Ref. 40, a numerical fit relating ${\sigma}_T^2(10^o)$ to $n$ and
${\sigma}_8$ can be found and equated to the observed $10^o$ variance
to yield
\be
\l{eq3.23}
{\sigma}_8=(1.18 \pm 0.2) e^{2.63(1-n)}{\sqrt{(3-n)/(15-13n)}}.
\ee
The contribution of the tensor modes arises solely in the rooted
factor and it becomes negligible as $n$ approaches unity. It has the
effect of {\em decreasing\/} ${\sigma}_8$ for a given spectral index,
thereby increasing the allowed bias.

A second result was presented in Ref. 39 based on the {\em
IRAS/QDOT\/}$^{13}$ and POTENT$^{43}$ galaxy surveys. Quoting a value of
$b_I=1.2 \pm 0.3$ at the 2-sigma level for the {\em IRAS\/} biasing factor,
these authors deduced that
\be
\l{eq3.24}
{\sigma}_8=(0.91 \pm 0.25) \left( {{1.9} \over {2.9-n}} \right)^{2/3}.
\ee
Eqs. (\ref{eq3.23}) and (\ref{eq3.24}) can be combined to yield a
bias-independent  lower limit on the spectral index of $n \ge 0.84$ for
consistency between  {\em COBE\/} and {\em QDOT\/}. It is this limit which
most strongly constrains the JBD parameter during inflation. If
Eq. (3.11) is valid in the pure JBD theory one recovers the
old extended inflationary solution.$^{44}$ In this regime the spectra have
the power law form discussed above with a spectral index
\be
\l{5.4}
n=1-\frac{8}{2\omega(\Psi)-1}.
\ee
Consistency with the CMBR and {\em QDOT} results therefore requires
$\omega> 25$ at the start of the last 60 e\---foldings before the end
of inflation. It is important to note that the constraint due to
bubble collisions, $\omega < 18$, does not apply here because the
phase transition is second-order.$^{19}$

\subsection{Nucleosynthesis and solar system constraints}

After the expansion becomes subluminal the most important constraints
on the value of $\omega$ and the vacuum energy arise at the
nucleosynthesis era and the current epoch respectively.  The standard
model of primordial nucleosynthesis can be used to constrain the
vacuum energy density and the JBD parameter at temperatures $\Ord(1)$
MeV. Time-varying gravitational and cosmological constants modify the
expansion rate of the universe and therefore the nuclear reaction
rates during this epoch. At temperatures exceeding the `freeze-out'
temperature $T_F \approx 0.8$ MeV the neutrons and protons are held in
local thermodynamic equilibrium by weak interactions.$^{15}$ In the
standard big bang picture the neutron-to-proton ratio at $T \ge T_F$
is determined by the Boltzmann factor $(n/p)=\exp (-Q/T)$, where
$Q=1.293$ $ \MeV$ is the n-p mass difference. The energy density of
the universe is then ${\rho}={\pi}^2g_{\rm eff}T^4/30$, where $g_{\rm
eff}$ is the number of relativistic degrees of freedom. The nucleons
fall out of equilibrium when the reaction rate equals the expansion
rate of the universe, \ie
\be
\l{eq4.8}
G_W^2T^5_F \approx H \propto {\sqrt{G{\rho}}}.
\ee
The temperature drops below $T_F$ and free neutrons undergo
${\beta}$-decay until the photodissociation of deuterium becomes
energetically unfavourable at some temperature $T_D$. At this point
synthesis of $He^4$ proceeds rapidly.

The abundance of $He^4$ depends crucially on the number of neutrons
present at $T_D$. There will always be two competing effects in any
modification to the standard picture. Increasing the freeze-out
temperature increases $(n/p)$ and one might expect the $He^4$
abundance to be correspondingly higher. However, in many cases it then
takes longer for the universe to cool to $T=T_D$.  More neutrons can
undergo ${\beta}$-decay and this effectively reduces the $He^4$
abundance.

When vacuum energy is introduced the second effect is dominant and the
observed $He^4$ abundance therefore leads to an upper limit on
${\Omega}_V$.$^{14}$ From detailed numerical calculations it was found
that this limit is
\be
\l{eq4.9}
{\Omega}_V[1\MeV] \le 0.08
\ee
for three neutrino species.$^{14}$

There is also a limit on the change in Newton's constant with time. In
the standard JBD theory the strength of gravity is higher at earlier
times and this increases $T_F$ as indicated by Eq. (\ref{eq4.8}). One
therefore expects an upper limit on ${\dot G}/G$ to exist or
equivalently a lower limit on ${\omega}$. Casas {\em et al.}$^{45}$
found ${\omega}>250$, but by dropping some of the simplifying
assumptions used to obtain this bound, Serna {\em et al.}$^{46}$ found
the weaker limit of ${\omega} > 50$. As a first approximation we shall
apply this weaker limit to the more general scalar-tensor theories
under consideration here, \ie
\be
\l{eq4.10}
{\omega}[{\Psi}, 1\MeV] \ge 50.
\ee
It should be emphasized that this limit is only strictly valid when
${\omega}$ is constant for all time. Although constraints
(\ref{eq4.9}) and (\ref{eq4.10}) may be weaker than those derivable
from more exact calculations, they are sufficient to severely limit
the scenario discussed here, as is shown by the numerical calculations
in section 4.

Finally, constraints on the magnitude and form of the JBD parameter
(\ref{eq2.4}) at the present epoch can be obtained from solar system
experiments by using the post-Newtonian approximation.  In this
analysis one considers the time independent spherically symmetric
metric around a point mass $m$, expanded as a series in the
gravitational potential $U=m/r$, \ie
\be
\l{eq4.11}
ds^2=(1-2{\alpha}U+2{\beta}U+...)dt^2+(1+2{\gamma}U)d{\bf x}^2,
\ee
where $d{\bf x}^2=dx_idx^i$ $(i=1,2,3)$. Current observational limits on the
Post-Newtonian parameters are
\be
\alpha = 1\pm 10^{-4}, \qquad
(\gamma+1)/2 = 1 \pm 10^{-3}, \qquad
(2+2\gamma-\beta)/3 =1\pm 10^{-2},
\ee
whereas general relativity predicts the values
${\alpha}={\beta}={\gamma}=1$.  Nordtvedt$^{16}$ analyzed the
generalized scalar-tensor theory (\ref{eq2.2}), finding that
\be
\l{eq4.12}
{\beta}=1+{{{\dd{\omega}/\dd{\Psi}} \over {(4+2{\omega})(3+2{\omega})^2}}},
\ee
and
\be
\l{eq4.13}
{\gamma} = {{1+{\omega}} \over {2 +{\omega}}} \qquad \Longrightarrow \qquad
{\omega}[{\psi}_0,3{\rm K}] > 500.
\ee
It is important to note that ${\omega} \rightarrow +\infty $ is a
necessary but not sufficient condition for the recovery of general
relativity at the present epoch. We also require from
Eq. (\ref{eq4.12}) that $(\dd{\omega}/\dd{\Psi})/{\omega}^3
\rightarrow 0$ as ${\omega}$ diverges. However, this condition is
always satisfied when the dilaton is located within the vicinity of a
turning point in $h(\psi)$, because
\be
\l{tp}
\frac{\dd \omega/ \dd \Psi}{\omega^3}= 4\frac{(\dd h / \dd \psi)^4}{h^3}
\left( 1- 2h \frac{\dd^2 h}{\dd\psi^2} \left( \frac{\dd h}{\dd \psi}
\right)^{-2}
\right).
\ee

To summarize, the most important limits are $\omega>25$ during
inflation, $\omega>50$ at nucleosynthesis and $\omega >500$ at the
present epoch.

\section{Numerical Results}
\def\theequation{\thesection.\arabic{equation}}
\setcounter{equation}{0}

\subsection{Initial Conditions}

A number of plausible models were considered by numerically
integrating the system of equations (\ref{infmotN})-(\ref{fredN}) plus
(\ref{rayN}).  The analysis does not depend strongly on the precise
form of $h(\psi)$.  One only requires that it be at least $C^2$
continuous, contain a turning point and be normalized to unity at this
point. In this sense the scenario we are investigating is rather
generic.  To ease the calculation we specified
\be
\l{5.1}
\Psi=h(\psi) = \sin^2 \beta\psi, \qquad \beta ={\rm constant},
\ee
because this leads to a simple form for $\omega(\Psi)$ given by
\be
\l{5.2}
\omega(\Psi)=\frac{1}{8\beta^2} \frac{1}{1-\Psi^2}.
\ee
Although there are no known particle physics models which lead
directly to Eq.  (\ref{5.1}), it has a simple analytical form which
can be viewed as an approximation to a more complete theory. Indeed,
for small $\Psi$ one can treat Eq. (\ref{5.2}) as a perturbation to
the JBD theory up to terms including $\Ord(\Psi^2)$. General
relativity is recovered as $\Psi \rightarrow 1$.  Following Linde, the
initial conditions were specified by treating the quantum boundary
(QB) as the most natural set of initial conditions for chaotic
inflation in scalar-tensor theories.$^{47}$ This boundary corresponds
to the hypersurface at the Planck density and represents the earliest
times at which initial conditions can be placed on classical
fields. In theory (\ref{eq2.1}) the Planck mass is related to the
dilaton by $\m^2(\psi)=h(\psi)$ so the QB is reached when $V(\sigma_i)
\approx h^2(\psi_i) = \m^4$. There is therefore a 1-jet family of
initial conditions relating the values of $\sigma$ and $\psi$.
Indeed, the probability that the universe can be created from nothing
via quantum tunneling is
\be
\l{5.3}
p \approx \exp (-3\m^4(\psi)/8V(\sigma)),
\ee
which is only high on the QB.$^{48}$

In the numerical calculations we employ this argument as a first
approximation.  To be more accurate, though, one should also consider
the effects of quantum fluctuations in the two fields. Once it has
started inflation always occurs at some point in the universe if the
change in the fields due to quantum fluctuations with wavelength
larger than $H^{-1}$ exceeds the classical change in the time interval
$H^{-1}$ due to the equations of motion.$^{47}$ An island universe
resembling our own (almost) flat Friedmann space-time may only form
when such a situation is reversed, and the point of equality is a more
accurate estimate of initial conditions.

The evolution of the quantity $\dd \ln a/ \dd \ln t \equiv Ht$ with
respect to the number of e-foldings, $N = \ln a$, was investigated
because this allows a number of interesting features to be shown
diagrammatically. For example, if the scale factor expands as a power
law, $a \propto t^p$, $Ht=p={\rm constant}$ and the graph is a
horizontal line. On the other hand, the exponential (de Sitter)
expansion may be written as $N=Ht+{\rm constant}$ and this is a
straight line of gradient $\pi/4$. The critical solution for inflation
is the Milne universe, $Ht=1$, and the strong energy condition is
violated above this line. The end of inflation can therefore be
defined as the point where the graph cuts the line $Ht=1$.

\subsection{Decaying Power Law Potentials}

Witten has shown that potentials of the form $V \propto
\sigma^{-\alpha}$, where $\alpha>0$, can arise when supersymmetry is
spontaneously broken.$^{49}$ The symmetry breaking occurs when the
potential has non-zero values. Therefore, one would have a natural
candidate for the inflaton if $\sigma$ was the field responsible for
breaking supersymmetry. In Einstein gravity the potential is too steep
at small $\sigma$ to lead to inflation, but at large $\sigma$ an
intermediate inflationary solution $a \propto \exp(Ft^q)$ is found,
where $F$ and $q<1$ are positive constants.$^{50}$

The question we address is whether inflation at small $\sigma$ is
possible in the scalar-tensor theory discussed above and the numerical
results are shown in Figure (1) for $\alpha=10$ and $\beta = 0.07$. A
power law expansion with $p=1/(8\beta^2 )+1/2$ arises when
$100<N<260$. In this region $\omega(\Psi)$ is approximately constant
$(\Psi^2 \ll 1)$ and the solution corresponds to that found in the JBD
theory when the inflaton is held constant. This is the old extended
inflationary solution $a \propto t^{\omega+1/2}$.$^{44}$ Hence the
known results for the form of the perturbation spectra derived when
Eq. (3.11) is valid can be employed in this analysis and
Eq. (\ref{eq3.15}) applies.  Consistency with the CMBR and {\em
QDOT\/} results therefore requires $\beta <0.07$.

\vspace{.2cm}
{\centerline{\bf Figures (1a) \& (1b)}}
\vspace{.2cm}

Unfortunately there is a graceful exit problem in this model. Figures
(1a) and (1b) show how the solution temporarily approaches the Milne
limit as $\dot{\sigma}^2$ increases and Einstein gravity is recovered
at $N \approx 310$. At this point the intermediate solution takes over
where $Ht=qN+{\rm constant}$. The gradient of this line is consistent
with the analytical result $\alpha=4(q^{-1}-1)$ derived by
Barrow.$^{50}$ (This provides a useful check for the numerical
calculations.)  Because the de Sitter solution is an attractor at
infinity in this case, it is not possible to obtain subluminal
expansion unless the reheating process is highly efficient. Rather
than redshifting to zero, the radiation density grows during the power
law phase due to the continual decay of $\sigma$. This is a generic
feature of these models, but is not sufficient to establish a
radiation-dominated phase as required by nucleosynthesis. Naively one
may think that $\dot{\sigma}^2$ would increase sufficiently fast as
$\Psi \rightarrow 1$, but it is too small relative to $\Psi$ during
inflation to contribute significantly to the reheating.  We considered
models up to $\alpha =100$ and found the same qualitative behaviour.

\subsection{Exponential potentials without dissipation}

We proceeded to investigate the exponential potential $V(\sigma)=V_0
\exp (-\lambda\k\sigma)$ where $\{V_0,\lambda\}$ are independent
constants and $\k^2 =8\pi$. In Einstein gravity this potential leads
to an attractor power law solution $a \propto t^{2/\lambda^2}$ when
$\lambda^2<6$ and inflation occurs if $0<\lambda^2<2$. The attractor
solution becomes $ a \propto t^{1/3}$ for all $\lambda^2>6$ and the
physical reason for this is as follows.  In the spatially flat
Friedmann model the $\lambda^2<6$ attractor arises because the kinetic
and potential energies of $\sigma$ redshift at the same rate. This is
the deep reason why inflation can never end in this model.  When
$\lambda^2>6$, however, the potential is too steep and the field
becomes effectively massless as $V \rightarrow 0$.$^{51}$

A number of new features arise when the Planck mass is allowed to
vary. We find that inflation is possible when $\lambda^2>6$ and that
the expansion becomes subluminal without any fine-tuning. As an
example, we discuss results for $\lambda^2=12$ and $\beta^2 =1/8$. In
Figure (2) the dissipation term $C_V$ is removed for all time. The
$p=\omega+1/2$ power law solution is again found, thereby implying
that the dilaton dominates the dynamics. The expansion does indeed
become subluminal and settles around the value $p=1/2$. This result is
independent of the value of $\beta$ and is understood by investigating
Eq. (\ref{dilaton}).  For $U=0$ and $\omega$ approximately constant,
this equation can be rewritten in the form
\be
\l{5.5}
\ddot{\Psi}+3H\dot{\Psi}+\frac{\partial V_{\rm eff}}{\partial\Psi} \approx 0,
\ee
where $V_{\rm eff}$ is an effective interaction potential defined by the
integrability condition
\be
\l{5.6}
\frac{\pdf V_{\rm eff}}{\pdf\Psi} \equiv -\frac{8\pi}{2\omega(\Psi)+3}T
\ee
and $T \equiv 4V-\dot{\sigma}^2$. Eq. (\ref{5.5}) resembles the
equation of motion for a minimally coupled field and we may view
$\Psi$ as evolving along its effective interaction potential.  During
the inflationary epoch, $V \gg
\dot{\sigma}^2 $ and $V_{\rm eff}$ has negative gradient.  Hence the values of
$\Psi$ and $\dot{\sigma}^2$ increase. This means that $T$, and therefore
$\pdf V_{\rm eff}/\pdf\Psi$,  eventually change sign as $4V -
\dot{\sigma}^2$  passes through zero, thus causing the dilaton to slow and
reverse its motion. Hence, there exists  a damped oscillatory behaviour around
$4V=\dot{\sigma}^2$ and some constant value of $\Psi$. The theory soon
resembles a rescaled version of general relativity with a traceless
energy-momentum tensor; the solution is therefore equivalent to the
radiation-dominated universe $a \propto t^{1/2}$.

\vspace{.2cm}
{\centerline{\bf Figures (2a) \& (2b)}}
\vspace{.2cm}

Hence, the effect of a varying Planck mass is to change the attractor
for steep exponential potentials from $p=1/3$ to $p=1/2$. The above
argument should apply to {\em any\/} steep potential for which $T<0$
in Einstein gravity.  Including an uncoupled radiation component will
not alter the argument since its energy-momentum is identically
traceless and does not affect $V_{\rm eff}$.

\subsection{Exponential potentials with dissipation}

In Figure (3) the dissipation term is included for all time.  Figure
(3c) shows the radiation density growing during inflation and a
sufficient reheating temperature for baryogenesis to proceed is easy
to obtain in this model.  Moreover, from Figures (3b) and (3d), it is
seen that the radiation begins to dominate once Einstein gravity is
recovered and the dissipation prevents the quantity
$4V-\dot{\sigma}^2$ from passing through zero. However, a new scaling
solution is found, where $\rho_{\sigma}/\rho_{\rm rad} = {\rm
constant}$, as is shown in the late time behaviour of Figures (3a) and
(3d).

\vspace{.2cm}
{\centerline{\bf Figures (3a) - (3d)}}
\vspace{.2cm}

This solution may be derived analytically for the special case where
$C_V$ is constant. We consider a model in which a particle species
$X$, with equation of state $p_X=(\gamma-1)\rho_X$ for some constant
$\gamma$, is coupled to the inflaton through an interaction $\rho_X
\equiv \Gamma_X\dot{\sigma}^2$, where $\Gamma_X$ is constant. The
Bianchi identity and Friedmann equation for this system are
\bea
\l{5.7}
(1+2\Gamma_X)\ddot{\sigma}+3(1+\gamma\Gamma_X)H\dot{\sigma}+V_{\sigma}=0 \\
\l{5.8}
3H^2=\k^2(\rho_V+\rho_X)=\k^2 \left[ \left( \frac{1}{2}+\Gamma_X \right)
\dot{\sigma}^2+V \right],
\eea
where $\rho_V$ represents the inflaton energy density and we assume
Einstein gravity holds. By differentiating Eq. (\ref{5.8}) with
respect to cosmic time and substituting Eq. (\ref{5.7}) for
$\ddot{\sigma}$ we arrive at the `momentum' equation
\be\l{5.9}
\dot{H}=-\frac{\k^2}{2} ( 1+\gamma\Gamma_X)\dot{\sigma}^2.
\ee
This implies that Eq. (\ref{5.8}) may be rewritten in the Hamilton-Jacobi
form by defining a new time coordinate
\be\l{5.10}
t = -\frac{\k^2}{2} ( 1+\gamma\Gamma_X) \int^{\sigma} \dd \, \sigma^{\prime}
\left( \frac{\dd H}{\dd \sigma^{\prime}}
\right)^{-1}.
\ee
We find
\be
\l{5.11}
3\k^2(1+\gamma\Gamma_X)^2H^2-2(1+2\Gamma_X) \left( \frac{\dd H}{\dd
\sigma} \right)^2=\k^4(1+\gamma\Gamma_X)^2V,
\ee
which yields the attractor solution
\be
\l{5.12}
H=\sqrt{A} \exp \left( - \lambda\k\sigma /2 \right), \qquad \lambda\k\sigma =
2\ln \left[ \frac{\lambda^2\sqrt{A}}{2(1+\gamma\Gamma_X)} t
\right],
\ee
where $A$ is a positive constant. Hence the scale factor grows as a power law
$a \propto
t^p$, where
\be
\l{5.13}
p \equiv \frac{2}{\lambda^2} (1+\gamma\Gamma_X).
\ee
(When $\Gamma_X \rightarrow 0$ we recover the standard power law
solution with $p=2/\lambda^2$). The contribution of $X$ to the matter
content of this universe may be expressed through the quantity
$\Omega_X \equiv
\k^2\rho_X/3H^2= \rho_X/(\rho_X+\rho_V)$. For the solution  (\ref{5.12}) we
find
\be
\l{5.14}
\Omega_X=\frac{\Gamma_X\lambda^2}{3(1+\gamma\Gamma_X)^2}
\ee
is constant, implying Eq. (\ref{5.12}) represents a scaling solution and
the matter and radiation  densities redshift as $\rho_V \propto \rho_X \propto
t^{-2} \propto a^{-2/p}$. Moreover, by substituting  Eq. (\ref{5.12}) into Eq.
(\ref{5.14}), the condition $V_0>0$ implies this attractor exists only if
\be
\l{5.15}
\Gamma_X>\frac{\Omega_X}{2(1-\Omega_X)}.
\ee
The nucleosynthesis constraint (\ref{eq4.9}) then implies  $\Gamma_X>23/4$
which
yields the lower limit  $\lambda^2 \ge 36$.

In principle this result suggests the nucleosynthesis constraints can
be satisfied in this model if $\lambda^2$ is sufficiently
large. However, if the dissipation is not removed, the scaling
solution will survive through to the decoupling era. Numerical results
have shown that the lack of observed spectral distortions in the CMBR
imply $\Omega_V < 4 \times 10^{-4}$ if the vacuum decays into low
energy photons.$^{14}$ For consistency, Eq. (\ref{5.15}) then implies
$\lambda^2 > \Ord(10^3)$, which is clearly unrealistic. In any case it
is physically reasonable to suppose the dissipation becomes negligible
once the effective inflaton mass falls below the rest mass of its
decay product.  Naively one would expect the vacuum energy to rapidly
redshift to zero at energies below this mass scale, with the radiation
density falling as $a^{-4}$. In general relativity this would be the
case, but the evolution of the Planck mass again significantly alters
the arguments, as shown in Figure (4).

\vspace{.2cm}
{\centerline{\bf Figures (4a) \& (4b)}}
\vspace{.2cm}

As soon as the dissipation term was removed the vacuum rapidly
dominated the universe once more. This feature arises because the peak
of reheating occurs while the dynamics is still dominated by the
dilaton's motion.  The dilaton viscosity still slows the inflaton and
thus its energy density does not decay as fast as it would in Einstein
gravity.  On the other hand, the time\---dependence of the Planck mass
does not affect the radiation, which still decays as $\rho_{r} \propto
a^{-4}$.  This follows because the energy\---momentum tensor of the
free radiation field is traceless and does not appear in the dilaton
field equation (2.12).  We suggest that this qualitative behaviour
should not depend too strongly on the precise form of the potential
unless $V(\sigma)$ is very steep. In this case, though, the vacuum
energy would be insignificant at the present epoch. The same
qualitative behaviour was observed for values of $\omega = 500$
$(\beta = 0.0158)$.

\vspace{1cm}
\section{Conclusions and Implications}
\vspace{1cm}

The philosophy behind this work was to identify a decaying
cosmological constant at the present epoch {\em directly} with the
vacuum energy which drove the inflationary expansion without altering
the form of the potential.  This would solve, without severe
fine\---tuning, the cosmological constant problem and could provide a
possible explanation for a number of observational results.  The
limiting solution for accelerated expansion, $a \propto t$, arises
when the vacuum energy density varies as $\rho_V \propto
a^{-2}$. Hence, $\rho_V$ must redshift faster than $a^{-2}$ if
inflation is to end, but must decay slower than the pressure\---free
matter component, $\rho_{\rm matter} \propto a^{-3}$, if it is to
dominate at late times. This requires viscosity to be present at early
times. We derived expressions for the amplitudes of scalar and tensor
fluctuations with a general inflaton potential and scalar\---tensor
theory, and the strongest constraints on any model arise from the
CMBR, primordial nucleosynthesis and solar system observations. These
constraints are important in any scenario of this type.  In general,
it is difficult (if not impossible) to satisfy these constraints
simultaneously if the viscosity arises due to a time\---dependent
Planck mass.

We considered two simple forms for a steep potential, $V \propto
\sigma^{-\alpha}$ and $V \propto \exp(-\lambda\sigma)$. In the former
a graceful exit problem arises for realistic values of $\alpha$. The
latter is more promising and a number of scaling solutions were found
both numerically and analytically.  In particular, inflation occurs
and ends naturally when $\lambda^2>2$. This model leads to a power
spectrum consistent with observation, but the vacuum does not decay
sufficiently fast to satisfy the nucleosynthesis constraint
(\ref{eq4.9}). It also appears that this constraint cannot be
satisfied for very steep potentials, which decay rapidly to zero in
Einstein gravity, because the Planck mass is still evolving during
reheating. This causes $\rho_V$ to redshift at a much reduced rate
relative to the radiation component.

To summarize, the scenario as outlined in section 2 is not viable for
the examples considered here and requires more complicated potentials
and further extensions. However, with this numerical code, it is
possible to develop working hyperextended chaotic models based on the
theories discussed in section 2 if the steep potential has a
minimum. One extension to this analysis is to investigate any effects
a dilaton self-interaction potential $U(\psi)$ may have. Alternatively
one could consider more general couplings of the dilaton and inflaton
fields in action (\ref{eq2.1}) or alter the form of the dissipation
term $C_V$ which models the reheating process.

These results may have implications for the joint evolution of the
comological and gravitational constants. A present\---day vacuum term
may arise if

1. The inflaton settles into a minimum of its potential at $V=0$ and a second
scalar
field is located in a  non\---zero minimum of its own potential.

2. The global minimum of the inflaton potential is located at $V \ne 0$.

3. $V(\sigma)$ decays monotonically for all time.

The third possibility is the most attractive, but based on the above
numerical calculations, one may conjecture that the vacuum will
generally not decay sufficiently fast relative to the relativistic
matter if $\m$ is time\---dependent during and shortly after
inflation. Our analysis therefore favours a reheating process via
oscillations of the inflaton about some minimum if the Planck mass
varies with time, but it does not rule out other possibilities. This
agrees indirectly, and for different reasons, with the conclusions of
Ref. 52, in which a detailed quantum mechanical description of the
reheating process was given. These authors concluded that particle
production is only significant during the oscillating phase of the
inflaton. (A direct comparison of conclusions cannot be made, however,
since these authors only considered reheating in Einstein gravity,
whereas the dilaton is still evolving during reheating in the models
discussed here.)

Conversely, if one prefers a continually decaying potential, this
suggests a constant Planck mass is required.  In any case, it appears
that some degree of fine\---tuning is necessary if a cosmological
constant arising from the inflaton potential is to be non\---zero at
the present epoch.

\vspace{1cm}

{\bf Acknowledgments} RAF is supported by a Science and Engineering
Research Council (S.E.R.C.), U. K., postgraduate studentship. JEL
receives an S.E.R.C. postdoctoral research fellowship and is supported
at Fermilab by the DOE and NASA under Grant NAGW-2381.  We thank
J. D. Barrow and P. Coles for useful discussions and comments and for
directing us to Ref. 16.

\newpage

{\bf References}

\vspace{1cm}

1. A. Einstein, {\em Sitz der Preuss Akad d Wiss} {\bf 1917}, 142 (1917).

2. F. Hoyle and J. V. Narlikar, {\em Proc Roy Soc} {\bf A273}, 1 (1963).

3. A. H. Guth, {\em Phys Rev} {\bf D23}, 347 (1981); K. A. Olive, {\em Phys
Rep} {\bf 190}, 307 (1990).

4. J. M. Bardeen, P. J. Steinhardt, and M. S. Turner, {\em Phys Rev} {\bf D28},
679 (1983).

5. A. Renzini and F. Pecci, {\em Ann Rev Astron Astro} {\bf 26}, 19 (1988);  M.

Rowan-Robinson, in {\em Observational tests of cosmological inflation} eds  T.
Shanks

{\em et al} (Kluwer Academic Publishers, London, 1991); D. Schramm, in {\em
Astrophysical

Ages and Dating Methods}, eds. E. Vangioni-Flam {\em et al.} (Gif-sur-Yvette,
France:

Editions Fronti\`eres)

p. 365; C. Delyannis {\em et al.} Yale University Preprint (1992).

6. E. W. Kolb and M. S. Turner, {\em The early universe}, (Addison-Wesley, New
York,

1990).

7. S. D. M. White, G. Efstathiou, and C. S. Frenk, Oxford preprint {\em The
amplitude

of mass fluctuations in the universe} (1993).

8. S. J. Maddox, G. Efstathiou, W. J. Sutherland, and J. Loveday, {\em Mon Not
Roy

Astron Soc} {\bf 242}, 43 (1990).

9. S. M. Carroll, W. H. Press, and E. L. Turner, {\em Ann Rev Astron
Astrophys}, {\bf 30}

499 (1992).

10. P. J. E. Peebles and B. Ratra, {\em Ap. J. Lett.} {\bf 325} L 17 (1988); B.
Ratra and P.

J. E. Peebles, {\em Phys Rev} {\bf D37}, 3406 (1988); B. Ratra, {\em Phys Rev}
{\bf D45}, 1913 (1992).

11. W. Saunders, {\em et al.}, {\em Nat} {\bf 349}, 32 (1991); and references
therein.

12. G. F. Smoot, {\em et al}, {\em Ap J Lett} {\bf 396}, L1 (1992).

13. W. Saunders, {\em et al.}, {\em Nat} {\bf 349}, 32 (1991); A. N. Taylor and
M. Rowan-Robinson,

{\em Nat} {\bf 359}, 396 (1992); M. Davies, F. J. Summers and D. Schlegel, {\em
Nat} {\bf 359}, 393

(1992).

14. K. Freese, F. C. Adams, J. A. Frieman, and E. Mottola, {\em Nucl Phys} {\bf
B287},

797 (1987).

15. K. A. Olive, D. N. Schramm, G. Steigman, and T. Walker, {\em Phys Lett}
{\bf B236},

454 (1990); T. Damour and C. Gundlach, {\em Phys Rev} {\bf D43}, 3873 (1991).

16. K. Nordtvedt, {\em Ap J} {\bf 161}, 1059 (1970): R. D. Reisenberg {\em et
al}, {\em Ap J Lett} {\bf 234},

L219 (1991).

17. P. G. Bergmann, {\em Int J Theo Phys} {\bf 1}, 25 (1968); R. V. Wagoner,
{\em Phys Rev} {\bf D

1}, 3204 (1970).

18. P. J. Steinhardt and F. S. Accetta, {\em Phys Rev Lett} {\bf 64}, 2740
(1990).

19. A. R. Liddle and D. Wands, {\em Mon Not Roy Astron Soc} {\bf 253}, 637
(1991); {\em Phys

Rev} {\bf D45}, 2665 (1992); J. D. Barrow and K. Maeda, {\em Nucl Phys} {\bf
B341}, 294 (1990).

20. T. Kaluza, {\em Preus Acad Wiss} {\bf 1}, 966 (1921); O. Klein, {\em Zeit
Phys} {\bf 37}, 895 (1926); {\em

Nat} {\bf 118}, 519 (1926); M. J. Duff, B. E. W. Nilsson, and C. P. Pope, {\em
Phys Rep}

{\bf 130}, 1 (1986).

21. R. Holman, E. W. Kolb, S. L. Vadas, and Y. Wang, {\em Phys Rev} {\bf D43}
995 (1991)

and references therein.

22. K. Maeda, {\em Phys Rev} {\bf D39}, 3159 (1989).

23. L. Amendola, M. Litterio, and F. Occhionero, {\em Phys Lett} {\bf B237},
348 (1990);

L. Amendola, S. Capozziello, M. Litterio, and F. Occhionero, {\em Phys Rev}
{\bf D45},

417 (1992).

24. B. A. Campbell, A. D. Linde, and K. A. Olive, {\em Nucl Phys} {\bf B355},
146 (1991).

25. A. B. Burd and A. Coley, {\em Phys Lett} {\bf B267}, 330 (1991).

26. M. Morikawa and M. Sasaki, {\em Prog Theo Phys} {\bf 72}, 782 (1984).

27. J Yokoyama and K. Maeda, {\em Phys Lett} {\bf B207}, 31 (1988).

28. A. L. Berkin and K. Maeda, {\em Phys Rev} {\bf D44}, 1691 (1991); N.
Deruelle, C.

Gundlach, and D. Polarski, {\em Class. Quantum Grav.} {\bf 9}, 137 (1992); S.
Mollerach

and S. Matarrese, {\em Phys. Rev.} {\bf D45}, 1961 (1992); N. Deruelle, G.
Gundlach,

and D. Langlois, {\em Phys. Rev.} {\bf D45}, R3301 (1992).

29. J. M. Bardeen, {\em Phys Rev} {\bf D22}, 1882 (1980).

30. J. E. Lidsey, {\em Class Quantum Grav} {\bf 9}, 149 (1992).

31. D. H. Lyth, {\em Phys Rev} {\bf D31}, 1792 (1985).

32. A. D. Linde, {\em Phys Lett} {\bf B108}, 389 (1982).

33. J. McDonald, {\em Phys Rev} {\bf D44}, 2314 (1991).

34. A. H. Guth and B. Jain, {\em Phys Rev} {\bf D45}, 441 (1992).

35. E. W. Kolb, D. S. Salopek, and M. S. Turner, {\em Phys Rev} {\bf D43}, 3925
(1991).

36. L. F. Abbott and M. B. Wise, {\em Nucl Phys} {\bf B244}, 541 (1984).

37. S. Kalara, N. Kaloper, and K. A. Olive, {\em Nucl Phys} {\bf B341}, 252
(1990).

38. F. Lucchin and S. Matarrese, {\em Phys Rev} {\bf D32}, 1316 (1985); {\em
Phys Lett} {\bf B164},

282 (1985).

39. A. R. Liddle and D. H. Lyth, {\em Phys Lett} {\bf B291}, 391 (1992); To be
published, {\em

Phys Rep} (1993).

40. F. C. Adams, J. R. Bond, K. Freese, J. A. Frieman, and A. V. Olinto, {\em
Phys

Rev} {\bf D47}, 426 (1993).

41. R. K. Sachs and A. M. Wolfe, {\em Ap J} {\bf 147}, 73 (1967).

42. J. E. Lidsey and P. Coles, {\em Mon Not Roy Astron Soc} {\bf 258}, 57P
(1992); R. L.

Davies, H. M. Hodges, G. F. Smoot, P. J. Steinhardt, and M. S. Turner, {\em
Phys Rev

Lett} {\bf 69}, 1851 (1992); L. M. Krauss and M. White, {\em Phys Rev Lett}
{\bf 69}, 869 (1992).

43. E. Bertschinger, A. Dekel, S. M. Faber, A. S. Dressler, and D. Burtstein,
{\em Ap

J} {\bf 354}, 370 (1990).

44. D. La and P. J. Steinhardt, {\em Phys Rev Lett} {\bf 62}, 376 (1989).

45. J. A. Casas, J. Garcia-Bellido, and M. Quiros, {\em Mod Phys Lett} {\bf
A7}, 447 (1992);

{\em Phys Lett} {\bf B278}, 94 (1992).

46.  A. Serna, R. Dom\'inguez-Tenreiro, and G. Yepes, {\em Ap J} {\bf 391}, 433
(1992).

47. A. D. Linde, {\em Phys Lett} {\bf B249}, 18 (1990).

48. A. Vilenkin, {\em Phys Rev} {\bf D37}, 888 (1988).

49. E. Witten, {\em Nucl Phys} {\bf B202}, 253 (1980); {\em Phys Lett} {\bf
B115}, 202 (1980).

50. J. D. Barrow, {\em Phys Lett} {\bf B235}, 40 (1990).

51. A. R. Liddle, {\em Phys Lett} {\bf B220}, 502 (1989).

52. L. Kofman, A. D. Linde, and A. Starobinsky, Preprint.

53. M. J. Geller and J. P. Huchra, {\em Science} {\bf 246}, 879 (1989).

54. J. E. Lidsey,  {\em Quantum Physics of the Universe}, Waseda, Japan,
(Vistas in

Astronomy, 1993).


\newpage

{\center{\bf Figure Captions}}

\vspace{1cm}

{\em Figure 1:} Numerical solutions for the potential $V(\sigma) =
10^{-10} {\sigma}^{-10}$, with $\beta = 0.07$ and initial conditions
$\dot{\sigma}=\dot{\Psi} =0$. The dissipation coefficient $f=1.0$ and
the rest mass of the decay product $X$ is $m_X = 10^9$ GeV. Figure
(1a) illustrates the evolution of the power index of the
solution. Power law and intermediate inflationary expansion is
observed, but a graceful exit from the inflationary regime is not
found. Figure (1b) illustrates the evolution of the $\omega$-parameter
in terms of ${\rm tan}^{-1} \omega(\Psi)$. The intermediate solution
takes over once $\omega$ diverges to infinity (\ie ${\rm tan}^{-1}
\omega \rightarrow +\pi /2$) and Einstein gravity is recovered.

\vspace{1cm}

{\em Figure 2:} Numerical solutions for the potential $V \propto \exp
(-\lambda\kappa\sigma)$ with $\lambda^2=12$, $\beta = 1/8$ and initial
conditions $\dot{\sigma}=\dot{\Psi} =0$. Dissipation effects have been
removed and $f=0$. Figure (2a) shows that inflation is possible and
ends naturally as the expansion becomes subluminal. The expansion
approaches the attractor $p=1/2$ as $\omega (\psi)$ settles to the
constant value in Figure (2b). A rescaled version of general
relativity is recovered but the Planck mass is smaller than the
currently observed value.

\vspace{1cm}

{\em Figure 3:} The same conditions as in Figure (2), but with $\beta
= 0.025$ and $ f=1.0$. The vertical dashed line measures where the
expansion becomes subluminal. Figure (3c) plots the evolution of the
radiation density $(\rho_r)$ and the inflaton energy density
$(\rho_{\sigma})$. Figure (3d) plots the ratio of the inflaton energy
density to the total energy density. The effects of dissipation are
included for all time. A new scaling solution is found once Einstein
gravity is recovered, but this violates spectral distortion limits on
the CMBR.

\vspace{1cm}

{\em Figure 4:} As Figure (3), but the dissipation terms are removed
at $m_X = 10^{10}$ GeV. In Figure (4a) the inflaton rapidly dominates
the expansion after the dissipation is removed. This follows because
the dilaton is evolving during the reheating, as shown in Figure (4b)
by the evolution of $\omega(\Psi)$.

\end{document}